\documentclass[aps,pra,twocolumn,showpacs,superscriptaddress,10pt,longbibliography]{revtex4-2}

\usepackage[colorlinks = true,linkcolor = red,citecolor = magenta]{hyperref}
\usepackage[sort&compress]{natbib}
\usepackage{scalerel}
\usepackage[normalem]{ulem}
\usepackage{xcolor}
\usepackage{bm}
\usepackage{amsmath}
\usepackage{amsfonts}
\usepackage{amssymb}
\usepackage{graphicx}
\usepackage{subfigure}
\usepackage{mathtools}
\usepackage{physics}
\usepackage{dsfont}
\usepackage{float}
\usepackage{lmodern}
\usepackage[many]{tcolorbox}
\usepackage{empheq}
\usepackage{cleveref}
\usepackage{textcomp}

\begin{document}

\title{Extraordinary resonant transmission  in two-terminal fermionic transport}
\author{P. S. Muraev}
\affiliation{IRC SQC, Siberian Federal University, 660041, Krasnoyarsk, Russia}
\author{D. N. Maksimov}
\affiliation{IRC SQC, Siberian Federal University, 660041, Krasnoyarsk, Russia}
\affiliation{Kirensky Institute of Physics, Federal Research Centre KSC SB RAS, 660036, Krasnoyarsk, Russia}
\author{A. R. Kolovsky}
\affiliation{IRC SQC, Siberian Federal University, 660041, Krasnoyarsk, Russia}
\affiliation{Kirensky Institute of Physics, Federal Research Centre KSC SB RAS, 660036, Krasnoyarsk, Russia}
\affiliation{School of Engineering Physics and Radio Electronics, Siberian Federal University, 660041, Krasnoyarsk, Russia}

\date{\today}

\begin{abstract}
We analyze conductance of a two-leg ladder connected with fermionic reservoirs, focusing on the decoherence effect induced by the reservoirs. In the absence of decoherence the system exhibits both bound states in continuum and Fano resonances. We found that the Fano resonances in transmittance are robust against decoherence, at the same time decoherence prevents collapse of resonances induced by bound states in continuum.
\end{abstract}
\maketitle

\begin{figure}
\includegraphics[width=8.5cm,clip]{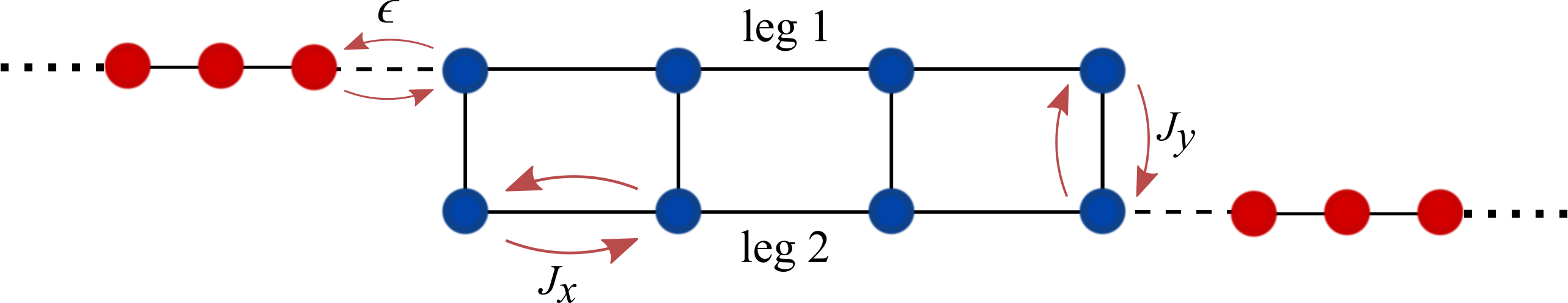}
\caption{Scattering problem for the two-leg ladder weakly ($\epsilon\ll J$) coupled to waveguides. Our control parameters will be the energy $E$ of the incoming wave and the ratio of the ladder hopping matrix elements $\xi=J_y/J_x$.}
\label{fig1}
\end{figure}

{\em 1.}
Resonant transmission is observed in a variety of physical systems, including solid-state mesoscopic devices. The effect of resonant transmissions for electrons follows from Landauer's theory which relates the system conductance to the transmission probability for the Bloch wave with a given Fermi quasimomentum\cite{Landauer1957,Datta1997}. Yet, in comparison with other systems such as, for example, photonic crystals \cite{Limonov17}, resonant transmission in solids has some peculiarities because there are no incident waves but rather fermionic reservoirs with differing chemical potentials that are connected to the device. The crucial property of any reservoir is its non-unitary relaxation dynamics which brings the reservoir to the thermal equilibrium. This relaxation dynamics causes partial decoherence of electron transporting states which, in its turn, modifies the resonant transmission \cite{131}. Until now, the decoherence effect of reservoirs has been studied only for systems with ordinary resonant transmission, where all transmission peaks have a Lorentzian shape. It is the aim of the present work to extend these studies onto the extraordinary resonant transmission. In this work, our model is a two-leg ladder shown in Fig.~\ref{fig1} which, along with the ordinary resonances, exhibits Fano resonances (FRs) and bound states in the continuum (BICs). 

 \vspace{0.5cm}
{\em 2.} Let us start with the tight-binding Hamiltonian of the two-leg ladder
\begin{equation}\label{sc_1}
    \widehat{\cal H} = -\frac{J_x}{2}\sum_{\ell=1}^{L-1}\Bigl(\hat{a}^{\dagger}_{\ell+1}\hat{a}_{\ell} + \hat{b}^{\dagger}_{\ell+1}\hat{b}_{\ell}\Bigl) -\frac{J_y}{2}\sum_{\ell=1}^{L}\hat{a}^{\dagger}_{\ell}\hat{b}_{\ell} + {\rm h.c.}
\end{equation}
where $\hat{a}^{\dagger}_{\ell}$ and $\hat{b}^{\dagger}_{\ell}$, $\hat{a}_{\ell}$ and $\hat{b}_{\ell}$ being fermionic creation and annihilation operators at the $\ell$th site in the 1st and 2nd leg of the ladder, $J_x$ is hopping constant between sites along the ladder and $J_y$ is the hopping constant between ladder legs. It is easy to show that the energy spectrum of the Hamiltonian \eqref{sc_1} is given by
\begin{equation}
\label{1}
E_{n} = -J_x\cos\left(\frac{\pi n}{L+1}\right) \mp \frac{J_y}{2}.
\end{equation}
and wave functions are
\begin{equation}
\label{2}
    \left[\begin{array}{c}
         \Psi_{\ell}(n)  \\
         \Tilde{\Psi}_{\ell}(n) 
    \end{array}\right] 
    = \frac{1}{\sqrt{L + 1}}\sin{\left(\frac{\pi n}{L + 1}\ell\right)}
    \left[\begin{array}{c}
         1  \\
         \pm 1 
    \end{array}\right] \;.
\end{equation}
Notice that wave functions can be sorted into two groups according to the symmetry $\Psi_{\ell}(n) =\pm \Tilde{\Psi}_{\ell}(L-n+1)$. 

\begin{figure}
\includegraphics[width=8.5cm,clip]{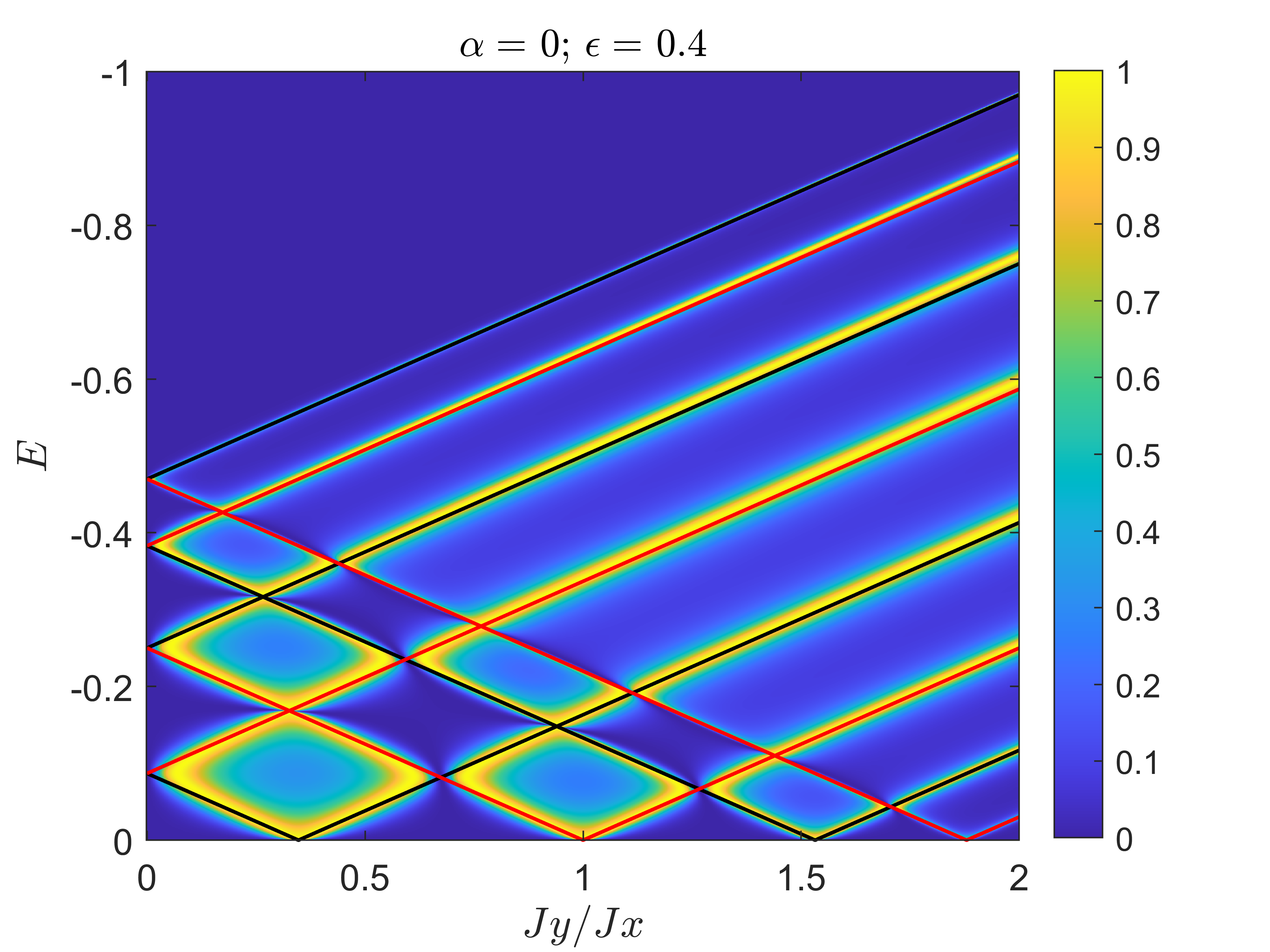}
\caption{Transmission probability  for the two-leg ladder as a color map for $\epsilon=0.4$. Superimposed are energy levels of the closed system where the black and red colors refer to odd and even symmetry of the wave functions.}
\label{fig2}
\end{figure}

Next, we discuss the transmission probability for the setup depicted in Fig.~\ref{fig1}. The relative simplicity of the system Eq.~\eqref{sc_1} allows us to find the scattering matrix of the open ladder exactly \cite{Sadreev2003, Maksimov2015},
\begin{eqnarray}
\label{3a} 
 {\cal S}_{j, j'} (E)= -\delta_{j, j'}
  - \frac{i\epsilon^2|\sin\kappa|}{J}  \; W^{\dagger}_j\frac{1}{\widehat{H}_{\rm eff} - E}W_{j'} \;,\\
\label{3b}  
  \widehat{H}_{\rm eff} = \widehat{H} - \frac{\epsilon ^2e^{i\kappa}}{2J}\sum_{j = {\rm L,R}}W_j W^{\dagger}_j \;.
\end{eqnarray}
Here index $j$ takes the values $j=L,R$,  $\kappa$ is the wave vector of the incoming plane wave with the energy $E=-J\cos \kappa$,
$\epsilon$ is the coupling constant, $\widehat{H}$ is the single-particle version of the Hamiltonian \eqref{sc_1}, and $W_j$ are the column vectors with one non-zero element that corresponds to the ladder site connected to the waveguides.  Using Eqs.~(\ref{3a})-(\ref{3b}) we calculate the transmission amplitude $t(E)=S_{L,R}(E)$ and plot the transmission probability $|t(E)|^2$ as a function of control parameters in Fig.~\ref{fig2}. Superimposed are the energy levels of the closed ladder with the black and red colors referring to the even and odd symmetry of the eigenfunctions. Of particular interest in this figure are points in the vicinity of level crossings (level crossing points) where one observes extraordinary resonant transmission.  Outside these regions one has the ordinary resonant transmission where the transmission probability is given by a sum of Lorentzians, 
\begin{equation}
\label{9a}
|t(E)|^2\approx \sum_{n=1}^L  \frac{\Gamma_n^2}{\Gamma_n^2+(E-E_n)^2} \;,  \quad 
\Gamma_n =\frac{\epsilon^2}{J}  |\Psi_1(n)\tilde{\Psi}_L(n)|  \;.
\end{equation}
We stress that Eq.~(\ref{9a}) implicitly  assumes that the widths of the neighboring resonances are less than the distance between them. Clearly, this condition is violated at the level crossing points which require a separate consideration.  

\begin{figure}
\includegraphics[width=8.5cm,clip]{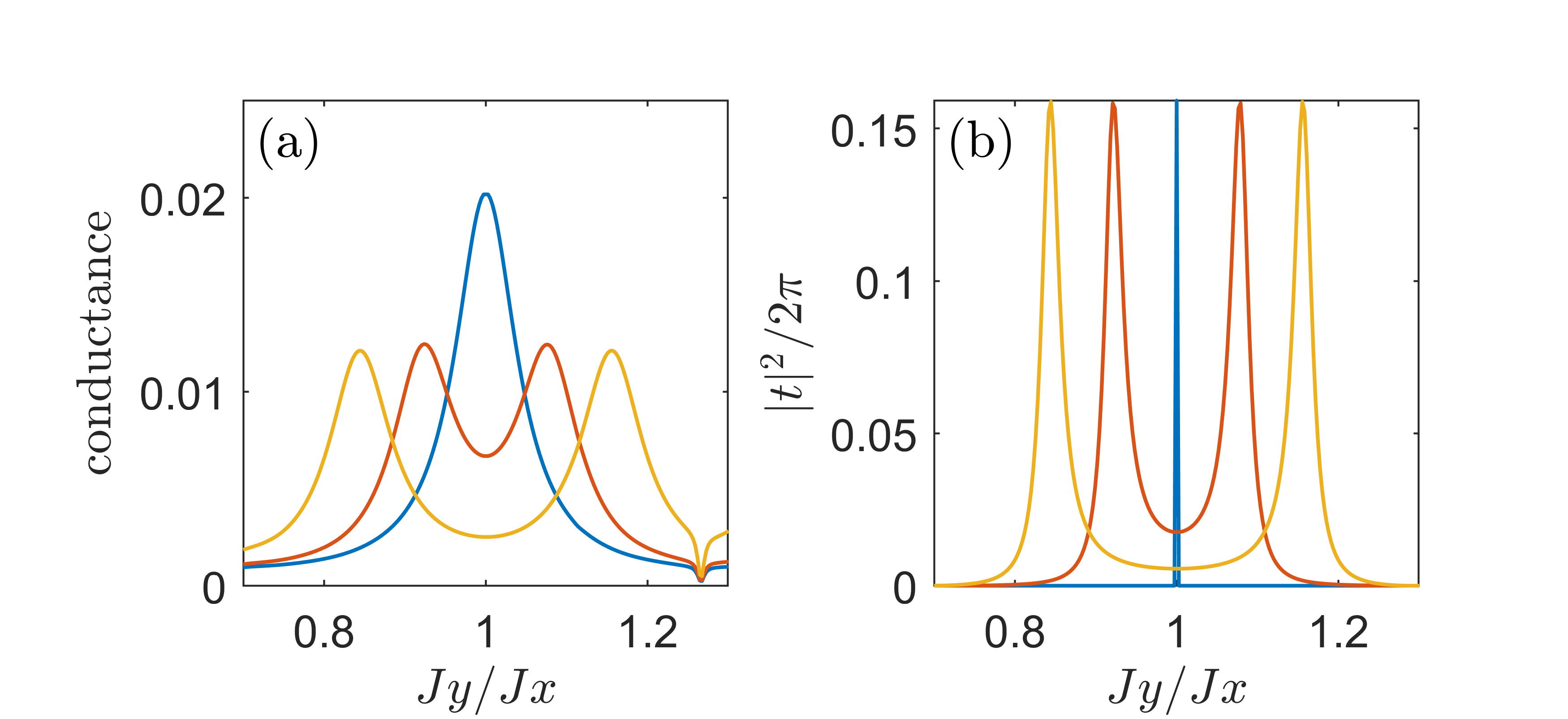}
\caption{Ladder conductance as the function of the control parameter $\xi$ for $\epsilon=0.2$ and the Fermi energy corresponding to BIC at $(E, \xi)=(0,1)$ (blue line),  and Fermi energies detuned from this level crossing point by $\Delta E=-0.02$ (red line), and $\Delta E=-0.04$ (yellow line). The left and right panels corresponds to $\gamma=0.02$ and $\gamma=0$, respectively.}
\label{fig3a}
\end{figure}

\begin{figure}
\includegraphics[width=8.5cm,clip]{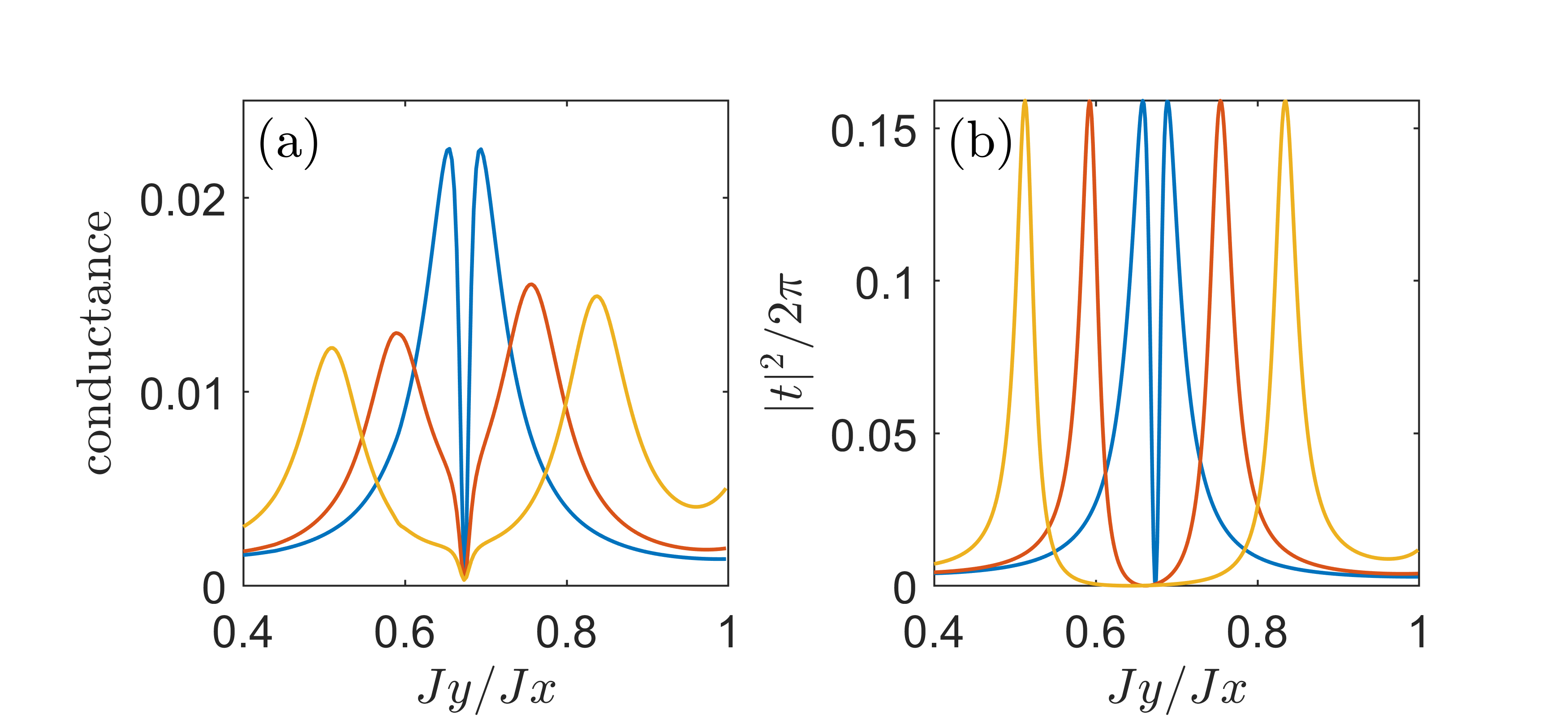}
\caption{The same as in Fig.~\ref{fig3a} yet for the Fermi energy corresponding to FR at $(E, \xi)=(-0.0816,0.6736)$.}
\label{fig3b}
\end{figure}

\begin{figure}
\includegraphics[width=8.5cm,clip]{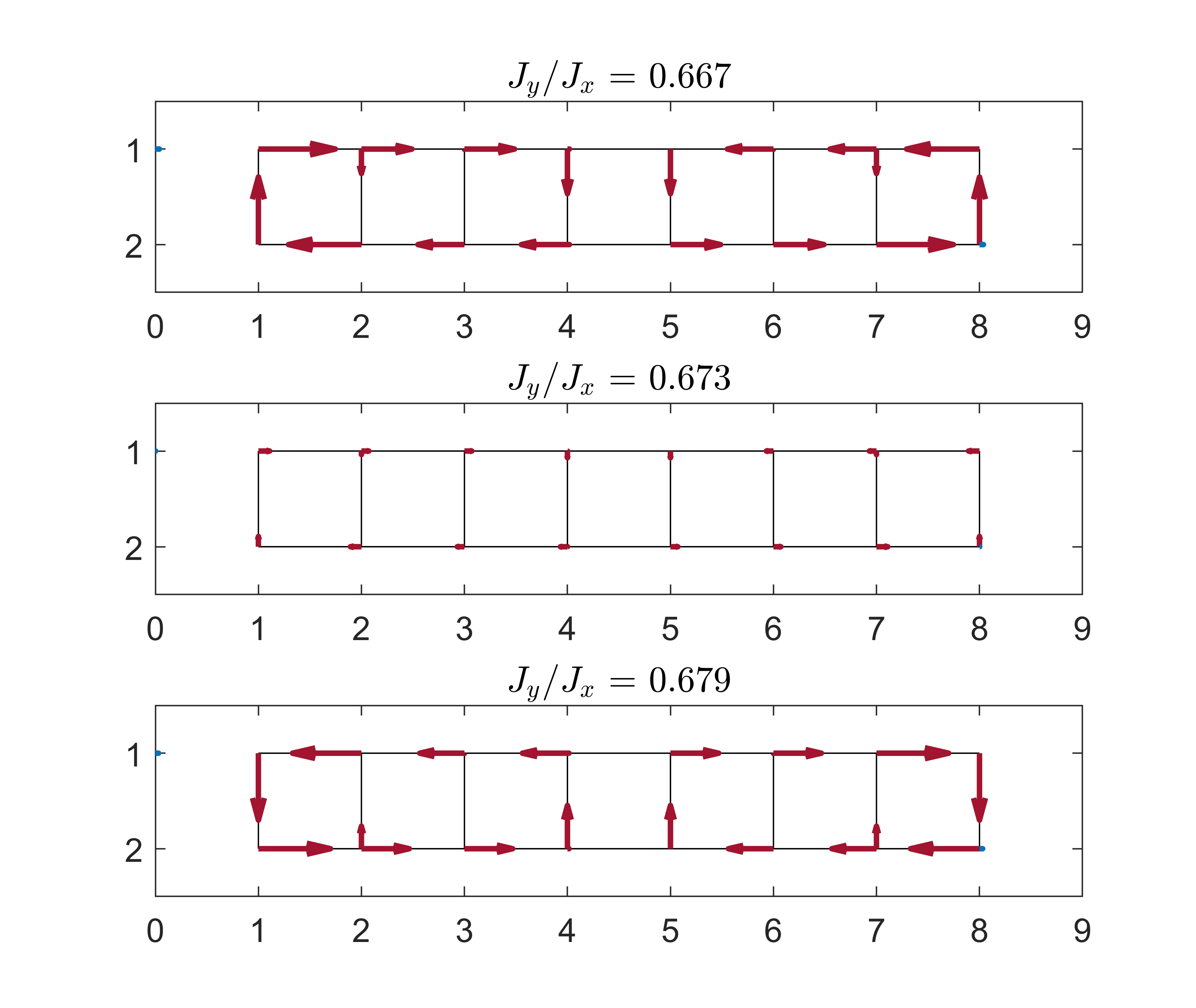}
\caption{The current pattern for FR depicted by blue line in in Fig.~\ref{fig3b}(b) in the vicinity of zero transmission point $\xi=0.673$. Notice that vortices  abruptly change their rotation directions at $\xi=0.673$. }
\label{fig5}
\end{figure}

To obtain a solution for the transmission amplitude in the vicinity of a level crossing point, we transform to the eigenbasis (\ref{2}) of the closed ladder and employ the two-mode approximation.  Within this approximation all matrix elements of the Hamiltonian (\ref{3b}) are set to zero except those associated with the two crossing levels. Then the four relevant matrix elements are
\begin{equation}
\label{5a}
    \widehat{H}^{\rm{(+)}}_{\rm eff} = \left[\begin{array}{cc}
         (E_m - e^{i\kappa}\Gamma_m) & - e^{i\kappa} \sqrt{\Gamma_n\Gamma_m} \\
         -e^{i\kappa} \sqrt{\Gamma_n\Gamma_m} & (E_n - e^{i\kappa}\Gamma_n)
    \end{array}\right] \;,
\end{equation}
if the levels belong to the same symmetry (the same color), and 
\begin{equation}
\label{5b}
    \widehat{H}^{\rm{(\pm)}}_{\rm eff} = \left[\begin{array}{cc}
         (E_m - e^{i\kappa}\Gamma_m) & 0 \\
         0 & (E_n - e^{i\kappa}\Gamma_n)
    \end{array}\right] 
\end{equation}
in the opposite case. After some algebra, we have
\begin{widetext}
\begin{equation}
\label{6a} 
    t^{\rm{(+)}} = -|\sin{\kappa}|\left(\frac{(E_n - E)\Gamma_n + (E_m - E)\Gamma_m}
    {(E_m - E)(E_n - E) - e^{i\kappa}(E_n - E)\Gamma_m - e^{i\kappa}(E_m - E)\Gamma_n}\right) \;,
\end{equation}
and
\begin{equation}
\label{6b} 
    t^{\rm{(\pm)}} = -|\sin{\kappa}|\left(\frac{\Gamma_m}{E_m - E - e^{i\kappa}\Gamma_m} 
    - \frac{\Gamma_n}{E_n - E - e^{i\kappa}\Gamma_n}\right) \;,
\end{equation}
\end{widetext}
Eqs.~(\ref{6a})-(\ref{6b}) describe two different scenarios of resonance merging. In the first case two resonances merge into the single resonance whose width shrinks to zero when we approach the level crossing point, see Fig.~\ref{fig3a}(b). In the second case, interacting resonances develop an avoided crossing  where the transmission amplitude is strictly zero at the level crossing point, see Fig.~\ref{fig3b}(b). Thus, we have either BIC or FR. We also mention that vanishing transmission does not imply that the system is empty of carriers. The origin of zero transmission lies in the specific current pattern. Namely, on approach the level crossing point the directed flow of probability changes to the vortex structure with two counter-rotating vortices, see Fig.~\ref{fig5}. This blocks the probability current through the ladder.
%

To conclude this section, we remark on the coupling geometry. Through the paper we consider the diagonal coupling, where the waveguides/leads are attached to the upper-left and lower-right corners of the ladder. Of course, one can also consider the geometries with waveguides attached either to the upper or to lower leg of the ladder. In the latter case, the transmission probability looks similar to that shown in Fig.~\ref{fig2}, but the BICs and the FRs swap their positions. Thus, one can study both BIC and FR focusing on the single point in the parameter space (for example, $E=0$ and $\xi=1$)  but changing the geometry of coupling, that might be useful from the experimental viewpoint. 

 \vspace{0.5cm}
{\em 3.} 
To study the effect of the external, i.e. induced by reservoirs, decoherence on the BICs and and the FRs we employ the model introduced in Refs.~\cite{Ajisaka2012, Gruss_2016, 120}. In what follows, we use a variant of the model from Refs.~\cite{120,131} where reservoirs are modeled by tight-binding rings of the size $M$,
\begin{equation}
\label{7a}
\widehat{{\cal H}}_{\rm r}=
-J\sum_{k=1}^{M}\cos\left(\frac{2\pi k}{M}\right) \hat{b}_{k}^{\dagger}\hat{b}_{k} \;,\\
\end{equation}
and the relaxation dynamics of reservoirs is mimicked by the Lindblad drain and gain operators,
\begin{eqnarray}
\label{7b}
\widehat{{\cal L}}^{(d)}_{j}=\gamma \sum_{k=1}^M\frac{\bar{n}_{k,j}-1}{2}
\left(\hat{b}_{k}^{\dagger}\hat{b}_{k}\widehat{\cal R }-2\hat{b}_{k}\widehat{\cal R }\hat{b}_{k}^{\dagger}
+\widehat{\cal R }\hat{b}_{k}^{\dagger}\hat{b}_{k} \right) ,\\
\nonumber
\widehat{{\cal L}}^{(g)}_{j}=-\gamma \sum_{k=1}^M\frac{\bar{n}_{k,j}}{2}
\left(\hat{b}_{k}\hat{b}_{k}^{\dagger}\widehat{\cal R }-2\hat{b}_{k}^{\dagger}\widehat{\cal R }\hat{b}_{k}
+\widehat{\cal R }\hat{b}_{k}\hat{b}_{k}^{\dagger} \right) ,
\end{eqnarray}
In Eqs.(\ref{7a})-(\ref{7b}) $\hat{b}^\dagger_k$ and $\hat{b}_k$ are the creation and annihilation operators which create/annihilate a fermion in the Bloch state with the quasimomentum $\kappa=2\pi k/M$, $\widehat{{\cal R}}$ is the density matrix of the whole system, i.e. the ladder and the rings, and $\gamma$ is the relaxation rate to the Fermi-Dirac distribution, 
\begin{equation}
\label{7c}
\bar{n}_{k,j}= \frac{1}{e^{-\beta_{j}[J\cos(2\pi k/M)+\mu_{j}]}+1}  \;,
\end{equation}
which is parametrized by the chemical potential $\mu_{j}$ and the temperature $T_j=1/\beta_j$ of the respective reservoir.  For the sake of simplicity, we shall consider zero temperature.  Then the chemical potential coincides with the Fermi energy. 

We find the stationary solution of the master equation for the density matrix $\widehat{{\cal R}}$,
\begin{equation}
\label{8a}
\frac{\partial \widehat{{\cal R}}}{\partial t}=-i[\widehat{{\cal H}}, \widehat{{\cal R}}]+
\sum_{j=L,R}\left(\widehat{{\cal L}}^{(g)}_{j}+\widehat{{\cal L}}^{(d)}_{j}\right) ,
\end{equation}
and calculate the current through the ladder depending on the chemical potential difference $\Delta\mu=\mu_{L}-\mu_{R}$.  Next, considering the limits $M\rightarrow\infty$ and $\Delta\mu\rightarrow0$ we calculate the conductance $\sigma=\sigma(E)$. The results obtained are depicted in Fig.~\ref{fig3a} and Fig.~\ref{fig3b} where they are compared with the Landauer's formula $\sigma(E)=G |t(E)|^2$ with $G=1/2\pi$ being the conductance quantum in the dimensionless units. It is seen that in the case of ordinary resonant transmission external decoherence broadens the resonant peaks \cite{131},
\begin{equation}
\label{9b}
\sigma(E)\approx \frac{1}{2\pi} \sum_{n=1}^L \frac{\Gamma_n(\Gamma_n+\gamma/2)}{(\Gamma_n+\gamma/2)^2+(E-E_n)^2} \;.\end{equation}
Equation~(\ref{9b}) interpolates between the case $\gamma=0$ where the system is perfectly conducting at $E=E_n$, and the case of large $\gamma\gg\Gamma_n$, where the system conductance at $E=E_n$ is inverse proportional to $\gamma$.

The effect of external decoherence on extraordinary resonant transmission is less trivial. In the BIC case, the crucial observation is that the blue curve is also well approximated  by Eq.~(\ref{9b}). Thus, the external decoherence prevents the collapse of resonances induced by a BIC. This is not the case with the FR. It is seen Fig.~\ref{fig3b}(a) that spectral feature, which leads to the Fano shape of the resonant peak a the level crossing point, survives external decoherence.

\begin{figure}
\includegraphics[width=8.5cm,clip]{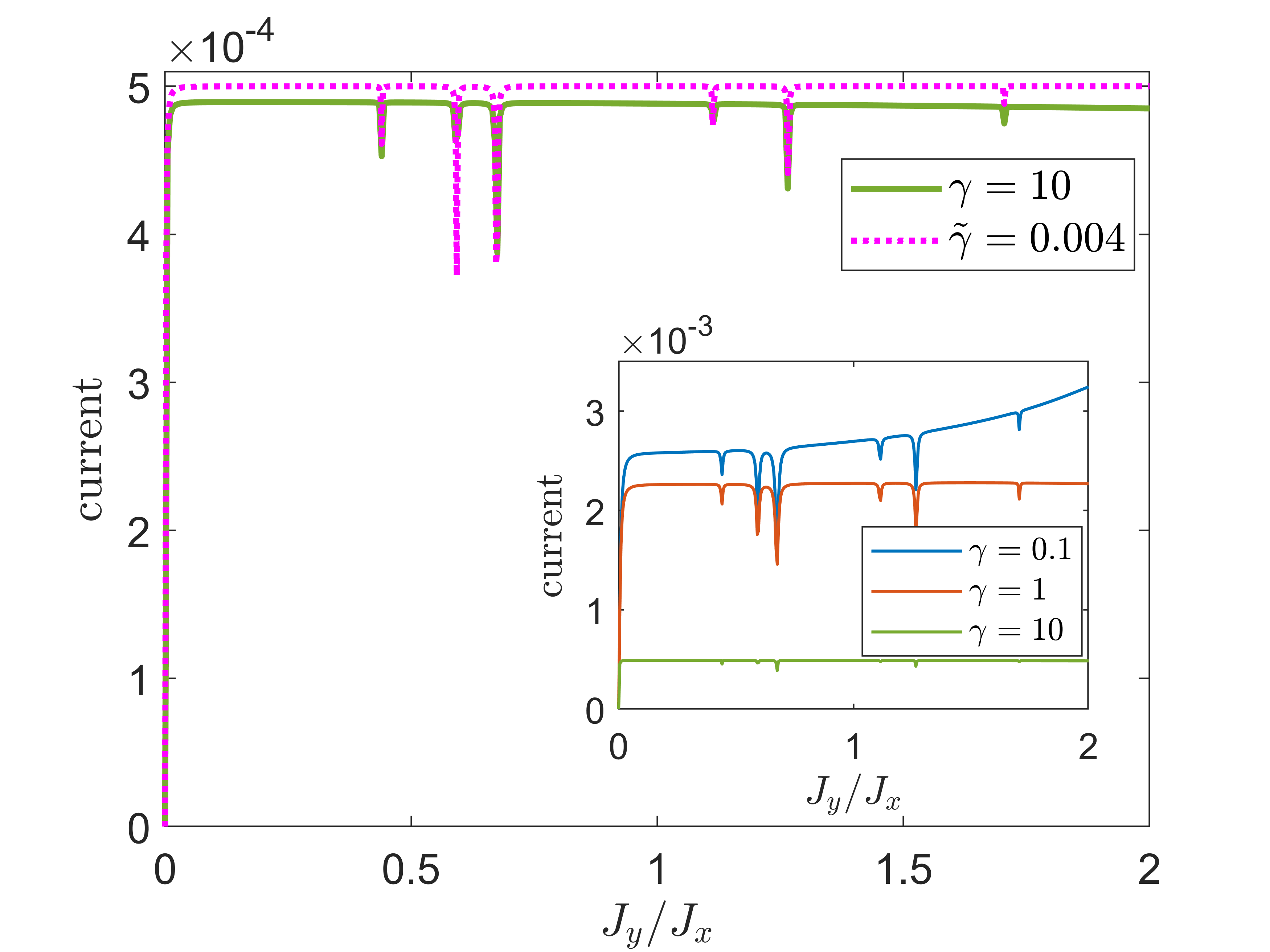}
\caption{The ladder conductance calculated on the basis of the original model for the parameters justifying the Born-Markov approximation: $\beta=1$, $\epsilon=0.2$, $\gamma=10$. The values of the chemical potentials correspond to the mean particle densities  $\bar{n}_{\rm L}=1$ and $\bar{n}_{\rm R}=0.5$ in the left and right rings, respectively. The dotted line is the result of Markovian master equation \eqref{11a} for $\tilde{\gamma}=\varepsilon^2/\gamma=0.004$. The inset shows convergence to the Markovian case as $\gamma$ is increased. }
\label{fig4}
\end{figure}
{\em 4.} Since the FRs are found to be resistant to a weak decoherence it is interesting to consider the limit of strong decoherence where the reservoir relaxation time $\tau \sim 1/\gamma$ is the smallest time scale of the problem. In this case, the reduced density matrix of the system, $\widehat{{\cal R}}_s = {\rm Tr}_r[\widehat{{\cal R}}]$, can be shown to obey the following Markovian master equation \cite{122}
\begin{equation}
\label{11a}
\frac{\partial \widehat{{\cal R}}_s}{\partial t}=-i[\widehat{{\cal H}}_s, \widehat{{\cal R}}_s]+
\sum_{{\ell}=L,R}\left(\widehat{{\cal L}}^{(g)}_{\ell}+\widehat{{\cal L}}^{(d)}_{\ell}\right) \;.
\end{equation}
In this equation the Lindblad drain and gain operators have the form
\begin{align}\label{11b}
\begin{split}
    \widehat{{\cal L}}^{(d)}_{\ell} &= \tilde{\gamma} \frac{\bar{n}_{\ell}-1}{2}
\left(\hat{c}_{\ell}^{\dagger}\hat{c}_{\ell}\widehat{\cal R }_s-2\hat{b}_{\ell}\widehat{\cal R}_s\hat{c}_{\ell}^{\dagger}
+\widehat{\cal R }_s\hat{c}_{\ell}^{\dagger}\hat{c}_{\ell} \right) ,\\
    \widehat{{\cal L}}^{(g)}_{\ell} &= -\tilde{\gamma} \frac{\bar{n}_{\ell}}{2}
\left(\hat{c}_{\ell}\hat{c}_{\ell}^{\dagger}\widehat{\cal R }_s-2\hat{c}_{\ell}^{\dagger}\widehat{\cal R }_s\hat{c}_{\ell}
+\widehat{\cal R }_s\hat{c}_{\ell}\hat{c}_{\ell}^{\dagger} \right) ,
\end{split}
\end{align}
%
where $\hat{c}_{\ell}^{\dagger}$ and $\hat{c}_{\ell}$ are the creation and annihilation operators that create/annihilate a fermion in the ladder site $\ell$ which is connected to the respective reservoir, $\tilde{\gamma}=\epsilon^2/\gamma$, and $\bar{n}_{L}$ and $\bar{n}_{R}$ are particle densities in the left and right reservoirs. Equation~{\eqref{11a} is known in physical literature  as the boundary driven Fermi-Hubbard model and it has analytical solutions in a number of important cases \footnote{We mention, in passing, that Eq.~\eqref{12a} also describes the case of bosonic carriers \cite{112, Ivanov2013} and is an analogue of the equation for the spin current in boundary driven integrable spin chains \cite{Landi22, Karevski2009,Znidaric2010}.}.  In particular, for the simple tight-binding chain the current across the chain is given by the following simple equation,
\begin{equation}
\label{12a}
\bar{j} \sim \frac{J_x \tilde{\gamma}}{J_x^2+\tilde{\gamma}^2}\frac{\bar{n}_L-\bar{n}_R}{2} \;.
\end{equation}
It is seen from the above equation that the current  is a monotonic function of the chain parameters and so is the chain conductance. Namely, in the high-temperature limit where the Fermi-Dirac distribution is almost flat we have
\begin{equation}
\label{12b}
\sigma=\beta \bar{n} \frac{J_x \tilde{\gamma}}{J_x^2+\tilde{\gamma}^2} \;.
\end{equation}
We calculated current of the two-leg ladder in the Markovian regime, see Fig.~\ref{fig4}. As expected, no resonant peaks that could be associated with ladder energy levels $E_n$ are visible.  However, one observes a number of deeps that are associated with the crossing of energy levels of different symmetry. Clearly, these deeps are remnants of the FRs which are symmetry protected against external decoherence. 
%

{\em 5.} We analyzed quantum transport of Fermi particles through a finite length two-leg ladder connected to particles reservoirs with slightly different chemical potentials. We focused on the extraordinary resonant transmission effects known in the scattering theory as Fano resonances and bound states in continuum (BIC). We showed that the unavoidable decoherence effect induced by the reservoirs destroy the signatures of the BICs and transform this extraordinary resonant transmission with collapsing Fano feature into ordinary one with the Lorentzian shapes of the resonant peaks. The Fano resonances, however, were found to be resistant to decoherence induced by reservoirs. From the viewpoint of practical applications this means that the ladder conductance can be changed from the maximally possible value to zero by small variation of the ladder parameters. 

\bibliography{mybib}

\begin{thebibliography}{16}%
\makeatletter
\providecommand \@ifxundefined [1]{%
 \@ifx{#1\undefined}
}%
\providecommand \@ifnum [1]{%
 \ifnum #1\expandafter \@firstoftwo
 \else \expandafter \@secondoftwo
 \fi
}%
\providecommand \@ifx [1]{%
 \ifx #1\expandafter \@firstoftwo
 \else \expandafter \@secondoftwo
 \fi
}%
\providecommand \natexlab [1]{#1}%
\providecommand \enquote  [1]{``#1''}%
\providecommand \bibnamefont  [1]{#1}%
\providecommand \bibfnamefont [1]{#1}%
\providecommand \citenamefont [1]{#1}%
\providecommand \href@noop [0]{\@secondoftwo}%
\providecommand \href [0]{\begingroup \@sanitize@url \@href}%
\providecommand \@href[1]{\@@startlink{#1}\@@href}%
\providecommand \@@href[1]{\endgroup#1\@@endlink}%
\providecommand \@sanitize@url [0]{\catcode `\\12\catcode `\$12\catcode
  `\&12\catcode `\#12\catcode `\^12\catcode `\_12\catcode `\%12\relax}%
\providecommand \@@startlink[1]{}%
\providecommand \@@endlink[0]{}%
\providecommand \url  [0]{\begingroup\@sanitize@url \@url }%
\providecommand \@url [1]{\endgroup\@href {#1}{\urlprefix }}%
\providecommand \urlprefix  [0]{URL }%
\providecommand \Eprint [0]{\href }%
\providecommand \doibase [0]{https://doi.org/}%
\providecommand \selectlanguage [0]{\@gobble}%
\providecommand \bibinfo  [0]{\@secondoftwo}%
\providecommand \bibfield  [0]{\@secondoftwo}%
\providecommand \translation [1]{[#1]}%
\providecommand \BibitemOpen [0]{}%
\providecommand \bibitemStop [0]{}%
\providecommand \bibitemNoStop [0]{.\EOS\space}%
\providecommand \EOS [0]{\spacefactor3000\relax}%
\providecommand \BibitemShut  [1]{\csname bibitem#1\endcsname}%
\let\auto@bib@innerbib\@empty
\bibitem [{\citenamefont {Landauer}(1957)}]{Landauer1957}%
  \BibitemOpen
  \bibfield  {author} {\bibinfo {author} {\bibfnamefont {R.}~\bibnamefont
  {Landauer}},\ }\bibfield  {title} {\bibinfo {title} {Spatial {Variation of
  Currents and Fields Due to Localized Scatterers in Metallic Conduction}},\
  }\href {https://doi.org/10.1147/rd.13.0223} {\bibfield  {journal} {\bibinfo
  {journal} {IBM Journal of Research and Development}\ }\textbf {\bibinfo
  {volume} {1}},\ \bibinfo {pages} {223} (\bibinfo {year} {1957})}\BibitemShut
  {NoStop}%
\bibitem [{\citenamefont {Datta}(1997)}]{Datta1997}%
  \BibitemOpen
  \bibfield  {author} {\bibinfo {author} {\bibfnamefont {S.}~\bibnamefont
  {Datta}},\ }\href@noop {} {\emph {\bibinfo {title} {Electronic transport in
  mesoscopic systems}}},\ \bibinfo {series} {Cambridge studies in semiconductor
  physics and microelectronic engineering}\ No.~\bibinfo {number} {3}\
  (\bibinfo  {publisher} {Cambridge university press},\ \bibinfo {address}
  {Cambridge [u.a.]},\ \bibinfo {year} {1997})\BibitemShut {NoStop}%
\bibitem [{\citenamefont {Limonov}\ \emph {et~al.}(2017)\citenamefont
  {Limonov}, \citenamefont {Rybin}, \citenamefont {Poddubny},\ and\
  \citenamefont {Kivshar}}]{Limonov17}%
  \BibitemOpen
  \bibfield  {author} {\bibinfo {author} {\bibfnamefont {M.~F.}\ \bibnamefont
  {Limonov}}, \bibinfo {author} {\bibfnamefont {M.~V.}\ \bibnamefont {Rybin}},
  \bibinfo {author} {\bibfnamefont {A.~N.}\ \bibnamefont {Poddubny}},\ and\
  \bibinfo {author} {\bibfnamefont {Y.~S.}\ \bibnamefont {Kivshar}},\
  }\bibfield  {title} {\bibinfo {title} {Fano resonances in photonics},\ }\href
  {https://doi.org/10.1038/nphoton.2017.142} {\bibfield  {journal} {\bibinfo
  {journal} {Nature Photonics}\ }\textbf {\bibinfo {volume} {11}},\ \bibinfo
  {pages} {543} (\bibinfo {year} {2017})}\BibitemShut {NoStop}%
\bibitem [{\citenamefont {Kolovsky}(2024)}]{131}%
  \BibitemOpen
  \bibfield  {author} {\bibinfo {author} {\bibfnamefont {A.~R.}\ \bibnamefont
  {Kolovsky}},\ }\bibfield  {title} {\bibinfo {title} {{Deriving Landauer’s}
  formula by using the master equation approach},\ }\href
  {https://doi.org/10.1209/0295-5075/ad56c3} {\bibfield  {journal} {\bibinfo
  {journal} {Europhysics Letters}\ }\textbf {\bibinfo {volume} {146}},\
  \bibinfo {pages} {61001} (\bibinfo {year} {2024})}\BibitemShut {NoStop}%
\bibitem [{\citenamefont {Sadreev}\ and\ \citenamefont
  {Rotter}(2003)}]{Sadreev2003}%
  \BibitemOpen
  \bibfield  {author} {\bibinfo {author} {\bibfnamefont {A.~F.}\ \bibnamefont
  {Sadreev}}\ and\ \bibinfo {author} {\bibfnamefont {I.}~\bibnamefont
  {Rotter}},\ }\bibfield  {title} {\bibinfo {title} {S-matrix theory for
  transmission through billiards in tight-binding approach},\ }\href
  {https://doi.org/10.1088/0305-4470/36/45/005} {\bibfield  {journal} {\bibinfo
   {journal} {Journal of Physics A: Mathematical and General}\ }\textbf
  {\bibinfo {volume} {36}},\ \bibinfo {pages} {11413} (\bibinfo {year}
  {2003})}\BibitemShut {NoStop}%
\bibitem [{\citenamefont {Maksimov}\ \emph {et~al.}(2015)\citenamefont
  {Maksimov}, \citenamefont {Sadreev}, \citenamefont {Lyapina},\ and\
  \citenamefont {Pilipchuk}}]{Maksimov2015}%
  \BibitemOpen
  \bibfield  {author} {\bibinfo {author} {\bibfnamefont {D.~N.}\ \bibnamefont
  {Maksimov}}, \bibinfo {author} {\bibfnamefont {A.~F.}\ \bibnamefont
  {Sadreev}}, \bibinfo {author} {\bibfnamefont {A.~A.}\ \bibnamefont
  {Lyapina}},\ and\ \bibinfo {author} {\bibfnamefont {A.~S.}\ \bibnamefont
  {Pilipchuk}},\ }\bibfield  {title} {\bibinfo {title} {Coupled mode theory for
  acoustic resonators},\ }\href
  {https://doi.org/10.1016/j.wavemoti.2015.02.003} {\bibfield  {journal}
  {\bibinfo  {journal} {Wave Motion}\ }\textbf {\bibinfo {volume} {56}},\
  \bibinfo {pages} {52} (\bibinfo {year} {2015})}\BibitemShut {NoStop}%
\bibitem [{\citenamefont {Ajisaka}\ \emph {et~al.}(2012)\citenamefont
  {Ajisaka}, \citenamefont {Barra}, \citenamefont {Mej{\'\i}a-Monasterio},\
  and\ \citenamefont {Prosen}}]{Ajisaka2012}%
  \BibitemOpen
  \bibfield  {author} {\bibinfo {author} {\bibfnamefont {S.}~\bibnamefont
  {Ajisaka}}, \bibinfo {author} {\bibfnamefont {F.}~\bibnamefont {Barra}},
  \bibinfo {author} {\bibfnamefont {C.}~\bibnamefont {Mej{\'\i}a-Monasterio}},\
  and\ \bibinfo {author} {\bibfnamefont {T.}~\bibnamefont {Prosen}},\
  }\bibfield  {title} {\bibinfo {title} {Nonequlibrium particle and energy
  currents in quantum chains connected to mesoscopic {Fermi} reservoirs},\
  }\href {https://doi.org/10.1103/PhysRevB.86.125111} {\bibfield  {journal}
  {\bibinfo  {journal} {Physical Review B}\ }\textbf {\bibinfo {volume} {86}},\
  \bibinfo {pages} {125111} (\bibinfo {year} {2012})}\BibitemShut {NoStop}%
\bibitem [{\citenamefont {Gruss}\ \emph {et~al.}(2016)\citenamefont {Gruss},
  \citenamefont {Velizhanin},\ and\ \citenamefont {Zwolak}}]{Gruss_2016}%
  \BibitemOpen
  \bibfield  {author} {\bibinfo {author} {\bibfnamefont {D.}~\bibnamefont
  {Gruss}}, \bibinfo {author} {\bibfnamefont {K.~A.}\ \bibnamefont
  {Velizhanin}},\ and\ \bibinfo {author} {\bibfnamefont {M.}~\bibnamefont
  {Zwolak}},\ }\bibfield  {title} {\bibinfo {title} {Landauer’s formula with
  finite-time relaxation: Kramers’ crossover in electronic transport},\
  }\bibfield  {journal} {\bibinfo  {journal} {Scientific Reports}\ }\textbf
  {\bibinfo {volume} {6}},\ \href {https://doi.org/10.1038/srep24514}
  {10.1038/srep24514} (\bibinfo {year} {2016})\BibitemShut {NoStop}%
\bibitem [{\citenamefont {Kolovsky}(2020)}]{120}%
  \BibitemOpen
  \bibfield  {author} {\bibinfo {author} {\bibfnamefont {A.~R.}\ \bibnamefont
  {Kolovsky}},\ }\bibfield  {title} {\bibinfo {title} {{Open Fermi-Hubbard
  model: Landauer’s }versus master equation approaches},\ }\href
  {https://doi.org/10.1103/PhysRevB.102.174310} {\bibfield  {journal} {\bibinfo
   {journal} {Physical Review B}\ }\textbf {\bibinfo {volume} {102}},\ \bibinfo
  {pages} {174310} (\bibinfo {year} {2020})}\BibitemShut {NoStop}%
\bibitem [{\citenamefont {Kolovsky}\ and\ \citenamefont
  {Maksimov}(2021)}]{122}%
  \BibitemOpen
  \bibfield  {author} {\bibinfo {author} {\bibfnamefont {A.~R.}\ \bibnamefont
  {Kolovsky}}\ and\ \bibinfo {author} {\bibfnamefont {D.~N.}\ \bibnamefont
  {Maksimov}},\ }\bibfield  {title} {\bibinfo {title} {Resonant transmission of
  fermionic carriers: {Comparison} between solid-state physics and quantum
  optics approaches},\ }\href {https://doi.org/10.1103/PhysRevB.104.115115}
  {\bibfield  {journal} {\bibinfo  {journal} {Physical Review B}\ }\textbf
  {\bibinfo {volume} {104}},\ \bibinfo {pages} {115115} (\bibinfo {year}
  {2021})}\BibitemShut {NoStop}%
\bibitem [{Note1()}]{Note1}%
  \BibitemOpen
  \bibinfo {note} {We mention, in passing, that Eq.~\protect \eqref {12a} also
  describes the case of bosonic carriers \cite {112, Ivanov2013} and is an
  analogue of the equation for the spin current in boundary driven integrable
  spin chains \cite {Landi22, Karevski2009,Znidaric2010}.}\BibitemShut {Stop}%
\bibitem [{\citenamefont {Kolovsky}\ \emph {et~al.}(2018)\citenamefont
  {Kolovsky}, \citenamefont {Denis},\ and\ \citenamefont {Wimberger}}]{112}%
  \BibitemOpen
  \bibfield  {author} {\bibinfo {author} {\bibfnamefont {A.~R.}\ \bibnamefont
  {Kolovsky}}, \bibinfo {author} {\bibfnamefont {Z.}~\bibnamefont {Denis}},\
  and\ \bibinfo {author} {\bibfnamefont {S.}~\bibnamefont {Wimberger}},\
  }\bibfield  {title} {\bibinfo {title} {{Landauer-Büttiker} equation for
  bosonic carriers},\ }\href {https://doi.org/10.1103/PhysRevA.98.043623}
  {\bibfield  {journal} {\bibinfo  {journal} {Physical Review A}\ }\textbf
  {\bibinfo {volume} {98}},\ \bibinfo {pages} {043623} (\bibinfo {year}
  {2018})}\BibitemShut {NoStop}%
\bibitem [{\citenamefont {Ivanov}\ \emph {et~al.}(2013)\citenamefont {Ivanov},
  \citenamefont {Kordas}, \citenamefont {Komnik},\ and\ \citenamefont
  {Wimberger}}]{Ivanov2013}%
  \BibitemOpen
  \bibfield  {author} {\bibinfo {author} {\bibfnamefont {A.}~\bibnamefont
  {Ivanov}}, \bibinfo {author} {\bibfnamefont {G.}~\bibnamefont {Kordas}},
  \bibinfo {author} {\bibfnamefont {A.}~\bibnamefont {Komnik}},\ and\ \bibinfo
  {author} {\bibfnamefont {S.}~\bibnamefont {Wimberger}},\ }\bibfield  {title}
  {\bibinfo {title} {Bosonic transport through a chain of quantum dots},\
  }\bibfield  {journal} {\bibinfo  {journal} {The European Physical Journal B}\
  }\textbf {\bibinfo {volume} {86}},\ \href
  {https://doi.org/10.1140/epjb/e2013-40417-4} {10.1140/epjb/e2013-40417-4}
  (\bibinfo {year} {2013})\BibitemShut {NoStop}%
\bibitem [{\citenamefont {Landi}\ \emph {et~al.}(2022)\citenamefont {Landi},
  \citenamefont {Poletti},\ and\ \citenamefont {Schaller}}]{Landi22}%
  \BibitemOpen
  \bibfield  {author} {\bibinfo {author} {\bibfnamefont {G.~T.}\ \bibnamefont
  {Landi}}, \bibinfo {author} {\bibfnamefont {D.}~\bibnamefont {Poletti}},\
  and\ \bibinfo {author} {\bibfnamefont {G.}~\bibnamefont {Schaller}},\
  }\bibfield  {title} {\bibinfo {title} {Nonequilibrium boundary-driven quantum
  systems: Models, methods, and properties},\ }\href
  {https://doi.org/10.1103/RevModPhys.94.045006} {\bibfield  {journal}
  {\bibinfo  {journal} {Reviews of Modern Physics}\ }\textbf {\bibinfo {volume}
  {94}},\ \bibinfo {pages} {045006} (\bibinfo {year} {2022})}\BibitemShut
  {NoStop}%
\bibitem [{\citenamefont {Karevski}\ and\ \citenamefont
  {Platini}(2009)}]{Karevski2009}%
  \BibitemOpen
  \bibfield  {author} {\bibinfo {author} {\bibfnamefont {D.}~\bibnamefont
  {Karevski}}\ and\ \bibinfo {author} {\bibfnamefont {T.}~\bibnamefont
  {Platini}},\ }\bibfield  {title} {\bibinfo {title} {{Quantum Nonequilibrium
  Steady States Induced by Repeated Interactions}},\ }\href
  {https://doi.org/10.1103/PhysRevLett.102.207207} {\bibfield  {journal}
  {\bibinfo  {journal} {Physical Review Letters}\ }\textbf {\bibinfo {volume}
  {102}},\ \bibinfo {pages} {207207} (\bibinfo {year} {2009})}\BibitemShut
  {NoStop}%
\bibitem [{\citenamefont {{\v{Z}}nidari{\v{c}}}(2010)}]{Znidaric2010}%
  \BibitemOpen
  \bibfield  {author} {\bibinfo {author} {\bibfnamefont {M.}~\bibnamefont
  {{\v{Z}}nidari{\v{c}}}},\ }\bibfield  {title} {\bibinfo {title} {{A matrix
  product solution for a nonequilibrium steady state of an XX chain}},\ }\href
  {https://doi.org/10.1088/1751-8113/43/41/415004} {\bibfield  {journal}
  {\bibinfo  {journal} {Journal of Physics A: Mathematical and Theoretical}\
  }\textbf {\bibinfo {volume} {43}},\ \bibinfo {pages} {415004} (\bibinfo
  {year} {2010})}\BibitemShut {NoStop}%
\end{thebibliography}%

\end{document}